\documentclass{article}
\usepackage{arxiv}
\usepackage[utf8]{inputenc} 
\usepackage[T1]{fontenc}    
\usepackage{setspace}
\usepackage{hyperref}       
\usepackage{url}            
\usepackage{booktabs}       
\usepackage{amsfonts}       
\usepackage{nicefrac}       
\usepackage{microtype}      
\usepackage{lipsum}		
\usepackage{graphicx}
\usepackage{natbib}
\usepackage{lineno}
\usepackage{dsfont}
\usepackage{doi}
\usepackage{amsthm,amsmath,amssymb}
\usepackage{mathtools}
\usepackage{graphicx,bm}
\usepackage{color}
\usepackage{lscape}
\usepackage{rotating}
\usepackage{setspace}
\usepackage{algorithm}
\usepackage{algpseudocode}
\usepackage{epigraph}

\newcommand{\bs}[1] {\bm{#1}}
\def \resp {$\mathcal{G}$\ }

\title{A Multinomial Canonical Decomposition Model, with emphasis on the analysis of Multivariate Binary data}

\author{ \href{https://orcid.org/0000-0001-7308-6210}{\includegraphics[scale=0.06]{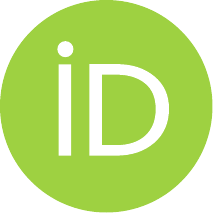}\hspace{1mm}Mark de Rooij}\\
 	Methodology and Statistics department, Leiden University\\
  	The Netherlands \\
  	\texttt{rooijm@fsw.leidenuniv.nl} \\
   }

\date{}

\renewcommand{\headeright}{}
\renewcommand{\undertitle}{}
\renewcommand{\shorttitle}{Canonical Decompositions of Categorical Data}

\begin{document}
\maketitle

\begin{abstract}
In this paper, we propose to decompose the canonical parameter of a multinomial model into a set of participant scores and category scores. External information about the participants or the categories can be used to restrict these scores. Therefore, we impose the constraint that the scores are linear combinations of the external variables. For the estimation of the parameters of the decomposition, we derive a majorization-minimization algorithm. We place special emphasis on the case where the categories represent profiles of binary response variables. In that case, the multinomial model becomes a regression model for multiple binary response variables and researchers might be interested in the effect of an external variable for the participant (i.e., a predictor) on a binary response variable or in the effect of this predictor on the association among binary response variables. We derive interpretational rules for these relationships in terms of changes in log odds or log odds ratios. Connections between our multinomial canonical decomposition and loglinear models, multinomial logistic regression, multinomial reduced rank logistic regression, and double constrained correspondence analysis are discussed. We use two empirical data sets, the first to show the relationships between a loglinear analysis approach and our modelling approach. The second data set is used as an illustration of our modelling approach and describes the model selection and interpretation in detail.    
\end{abstract}


\section*{Declarations}
Declarations of interest: none

\newpage

\section{Introduction}

In this paper, we are interested in the statistical analysis of a categorical variable $\mathcal{G} \in \{1, \ldots, K\}$, where $K$ might be very large. Participants ``choose'' or ``are in'' or ``belong to'' one of the classes or categories. The categorical variable \resp can be coded in an indicator matrix $\bm{G}$ with elements $g_{ik} \in \{0,1\}$ (for $i = 1,\ldots,N$ and $k = 1,\ldots,K$) where $g_{ik}$ indicates whether participant $i$ is in category $k$ ($g_{ik} = 1$) or not  ($g_{ik} = 0$), such that 
$\sum_{k=1}^K g_{ik} = 1$. 

Usually, the participants are described by a set of variables, for example demographic variables such as gender, age, race, education, etc. For the statistical analysis of a categorical outcome variable and a set of predictor variables a standard approach is multinomial logistic regression \citep{agresti2013categorical}. In some cases, however, additional information is also present for the categories of $\mathcal{G}$. \cite{harshman1982model}, for example, collected data from a large sample of Americans who bought a car. The outcome variable is the type of car the participant bought, which is a categorical variable with 106 different classes. Contemporary examples would include the Ford Mustang, Audi A4, Fiat 500, or Nissan Qashqai. The cars can be described by certain characteristics, like size, price, origin (America, Europe, Asia, ..), engine type (Gasoline, Diesel, Hybrid, or Electric), etc. A typical question for this type of data might be whether old white men buy more often Asian gasoline cars. This additional information about the categories cannot be included in a standard multinomial logistic regression.  

Another example of a categorical variable with additional information occurs when \resp is a profile of underlying variables. In that case, for every participant data is collected for $R$ response variables, $Y_1$, $Y_2$, ..., $Y_R$. Any combination of the categories of the response variables is called a profile. In this case, $K = \prod_{r=1}^R C_r$, where $C_r$ is the number of categories of response variable $r$ ($r = 1,\ldots,R$) and $K = 2^R$ when all response variables are dichotomous. $K$ grows quickly with the number of response variables, that is, with only 8 dichotomous response variables the number of categories is already $K = 256$ while with 11 dichotomous response variables $K = 2048$. Of course, $K$ also grows quickly when a few $C_r$ are large. In this case, the additional information represents the underlying response variables and their interactions.  

\cite{derooijkroonenberg2003}, for example, investigated the behavior of child-therapist pairs. A child and therapist are observed during a therapeutic session of approximately an hour, and their behaviors are scored every five seconds, where both the behavior of the therapist and of the child are coded using four categories: non play, play preparation, functional play, and imagery play. The cross-classification of the child's behaviour categories and the therapist's behaviour categories yields a categorical variable with 16 categories. This would be our major outcome variable, with the underlying child and therapist behavior as characteristics of the 16 categories. 

Yet another example is discussed in Agresti (2013, Table 10.1) where data is presented of senior high school students and their use of alcohol (A), cigarettes (C), and marijuana (M), all binary yes or no. This gives rise to multinomial response variable with 8 categories, representing the profile of alcohol, cigarettes and marijuana use. The three individual binary variables characterize the eight categories. In the data set, the students are described by their gender and race, but other individual characteristics like age could be included.   

For the analysis of categorical data there is a long history of statistical approaches \citep[see][chapter 17 for a historical tour]{agresti2013categorical}. Loglinear models \citep{bishop1975discrete, wickens1989multiway, agresti2013categorical} have been developed for the analysis of categorical data which are represented in contingency tables, that are, multi-way cross classifications of the variables with the number of occurrences of any combination in the cells. Loglinear models are a special case of generalized linear models \citep{McCullagh1989GLM} where the count variable acts a the response and the variables as predictors in a Poisson regression model. For the data as described above, loglinear models, however, cannot be applied when a characteristic of either the participant (row object) or response category (column object) is a numerical variable. 

Another approach for analyzing a categorical response variable is multinomial logistic regression \citep[][chapter 7]{agresti2013categorical}. In this model, the response variable (\resp) is explained by the predictor variables of the participants. The model results in $K -1$ regression equations, which might be daunting when $K$ is large. For ordinal multinomial responses, \cite{anderson1984regression} proposed a reduced rank multinomial model, called the stereotype model, where a rank 1 constraint is imposed on the regression coefficient matrix. In the same paper,  \cite{anderson1984regression} also discussed higher rank models, but these have received less attention. Reduced rank multinomial regression has also been described by \cite{yee2015}. In both approaches, the multinomial logistic regression model and its reduced rank variants, the characteristics of the classes, however, cannot be taken into account. 

In this paper, we develop a canonical decomposition model for analysing a multinomial response variable with potentially many categories, where we have additional information for both the participants  and the categories. In the next section, we will develop this model, derive an algorithm for estimation of the model parameters, and discuss model selection issues. In Section \ref{sec:dichotomous}, we describe in detail the case that the response variable represents a profile of binary responses. Interpretational rules will be derived that link the characteristics of the participants to the underlying binary responses and their associations. In Section \ref{sec:relations}, we describe relationships of our new approach to loglinear analysis, multinomial regression, reduced rank multinomial regression, and double constrained correspondence analysis. We apply our model, in Section \ref{sec:applications}, to two empirical data sets. We conclude this paper with a discussion.

\section{Canonical decomposition of a categorical variable}\label{sec:esm}

\subsection{Model}

Let us first define the categorical response variable \resp with $K$ categories or classes. The classes may simply be different types, such as the 106 different cars mentioned in the introduction, or may represent the response profile of a participant (see Section \ref{sec:dichotomous} for more details on this case). We assume a multinomial distribution for the variable \resp. Let us denote by $\pi_{ik}$ the probability that object $i$ ($i = 1,\ldots, N$) belongs to class or profile $k$ ($k = 1,\ldots,K$), that is, $\pi_{ik} = P(\mathcal{G}_i = k)$. A canonical decomposition model of this probability is given by 
\[
\pi_{ik} = \frac{\exp(\theta_{ik})}{\sum_{k' = 1}^K \exp(\theta_{ik'})}, 
\]
where $\theta_{ik}$ are the canonical parameters that are decomposed as
\[
\theta_{ik} = m_k + \bm{u}'_i\bm{v}_k,
\]
with $m_k$ representing an intercept for class $k$ and $\bm{u}_i$ and $\bm{v}_k$ vectors of length $S$, a number to be chosen by the user. The vector $\bm{u}_i$ represents \emph{object scores}, whereas the $\bm{v}_k$ represents \emph{class} or \emph{profile scores}. The number $S$ represents the rank or \emph{dimensionality} of the decomposition. The object scores $\bm{u}_i$ are collected in a matrix, that is
\[
\bm{U} = \left[ \bm{u}_1, \ldots, \bm{u}_N \right]'
\]
and, similarly, the profile scores $\bm{v}_k$ are collected in the matrix $\bm{V}$. 

For the participants or objects we usually have additional information. The user defined design matrix $\bm{X}$, of size $N \times P$, reflects this information. We assume the matrix $\bm{X}$ to be of full column rank. Having this information, we may require
\[
\bm{U} = \bm{XB}_x,
\]
that is, the object scores are a linear combination of the predictor variables $\bm{X}$. The matrix $\bm{B}_x$ are parameters to be estimated. 

For the classes or profiles we also have additional information. This information is collected in the design matrix $\bm{Z}$ of size $K \times Q$, again of full column rank. With this design matrix, we may require
\[
\bm{V} = \bm{ZB}_z,
\]
where the matrix $\bm{B}_z$ are parameters to be estimated. 

Finally, the additional information about the classes might also be used to constrain the intercepts. Therefore, collect the intercepts in the vector $\bm{m} = [m_1, \ldots, m_K]'$ and constrain this vector as
\[
\bm{m} = \bm{W}\bm{b}_w.
\]
The matrix $\bm{W}$ is of size $K \times T$ and the vector $\bm{b}_w$ is a vector of parameters. This constraint will be especially useful when the response variable \resp represents a profile. In that case, the parameters $\bm{m}$ model the associations among the response variables that create the profile and we can impose a simple structure on these associations by defining $\bm{W}$. In general, statistical models should be \emph{hierarchical} such that the lower order effects are present in a model whenever a higher order association is present. To fulfil that requirement, the columns of $\bm{Z}$ should be a subset of the columns of $\bm{W}$. 

Summarizing, the canonical parameters of our model are decomposed as 
\begin{eqnarray}
\theta_{ik} = \bm{w}'_k\bm{b}_w + \bm{x}'_i \bm{B}_x\bm{B}'_z \bm{z}_k, 
\label{eq:canonicaldecomp}
\end{eqnarray}
where $\bm{w}_k$, $\bm{x}_i$, and $\bm{z}_k$ are known design vectors. These vectors are rows of general design matrices $\bm{X}$, $\bm{Z}$, and $\bm{W}$. The design matrix $\bm{X}$ resembles design matrices for linear and logistic regression models, and in our decomposition model specifies which predictors or combination of predictors influences the responses. The design matrices $\bm{Z}$ and $\bm{W}$ resemble those of loglinear models. By defining $\bm{Z}$ we specify which (combinations of) characteristics is effected
by the predictors. So, $\bm{X}$ and $\bm{Z}$ together define which  predictors or interactions of predictors ($\bm{X}$) have an effect on which responses or associations among responses ($\bm{Z}$). The definition of $\bm{W}$ specifies which associations among the characteristics of the classes are included in the model. 

The probability form of the multinomial canonical decomposition model is
\[
\pi_{ik} = \frac{\exp(\bm{w}'_k\bm{b}_w + \bm{x}'_i \bm{B}_x\bm{B}'_z \bm{z}_k)}{\sum_{k' = 1}^K \exp(\bm{w}'_{k'}\bm{b}_w + \bm{x}'_i \bm{B}_x\bm{B}'_z \bm{z}_{k'})}. , 
\]
The parameters of the model are $\bm{b}_w$, $\bm{B}_x$ and $\bm{B}_z$. The next section shows how to estimate those parameters. 

\subsection{Maximum likelihood estimation} 

\subsubsection{Identification Constraints} 

The model as outlined in the previous section is not identified. The term
\[
\bm{UV}' = \bm{XB}_x\bm{B}'_z \bm{Z}'
\]
is a bilinear term and therefore
\[
\bm{UV}' = \bm{UT}\bm{T}^{-1}\bm{V}' = \bm{XB}_x\bm{T}\bm{T}^{-1}\bm{B}'_z \bm{Z}' = \bm{XB}^*_x\bm{B}^{*'}_z \bm{Z}' =\bm{U}^*\bm{V}^{*'}, 
\]
for any invertible matrix $\bm{T}$. We see that different sets of parameters give the same decomposition. To identify the solution, we require $\bm{V}'\bm{V} = \bm{I}$, that is, the matrix with profile scores should be orthonormal. Furthermore, we require the matrix $\bm{U}$ to be orthogonal, such that $\bm{U}'\bm{U}$ is a diagonal matrix. Using this constraints it is still possible that the sign of a column of $\bm{U}$ and $\bm{V}$ changes. Such sign switching can be prevented by choosing the first row of, for example, $\bm{V}$ to only have positive elements. 

Furthermore, adding a constant to the vector of intercepts $\bm{m}$ does not change the probabilities. Therefore, we require them to have zero mean or we will not include an intercept term in the design matrix $\bm{W}$.

\subsubsection{MM algorithm} 

In this section, we present a Majorization-Minimization (MM) algorithm \citep{heiser1995convergent, hunter2004tutorial, nguyen2017introduction} for the estimation of the model parameters. 
The idea of MM for finding a minimum of the function \(\mathcal{L}(\bs{\theta})\), where
\(\bs{\theta}\) is a vector of (canonical) parameters, is to define an auxiliary function, called a \emph{majorization function}, 
\(\mathcal{M}(\bs{\theta}|\bs{\vartheta})\) with two characteristics
\[
\mathcal{L}(\bs{\vartheta}) = \mathcal{M}(\bs{\vartheta}|\bs{\vartheta})\\
\] 
and 
\[
\mathcal{L}(\bs{\theta}) \leq \mathcal{M}(\bs{\theta}|\bs{\vartheta}),
\] 
where \(\bs{\vartheta}\) is a supporting point. The two equations tell us that \(\mathcal{M}(\bs{\theta}|\bs{\vartheta})\) is a function that lies above (i.e., majorizes) the original function and touches the original function at the support point. The support point at any iteration of the MM algorithm is defined by the current estimates of the (canonical) parameters.

A convergent algorithm can be constructed because
\[
\mathcal{L}(\bs{\theta}^+) \leq \mathcal{M}(\bs{\theta}^+|\bs{\vartheta}) \leq \mathcal{M}(\bs{\vartheta}|\bs{\vartheta}) = \mathcal{L}(\bs{\vartheta}),
\] 
where \(\bs{\theta}^+\) is 
\[
\bs{\theta}^+ = \mathrm{argmin}_{\bs{\theta}} \ \mathcal{M}(\bs{\theta}|\bs{\vartheta}),
\] 
the updated parameters. An advantage of MM algorithms is that they always converge monotonically to a (local) minimum. The challenge is to find a parametrized function family, \(\mathcal{M}(\bs{\theta}|\bs{\vartheta})\), that can be used in every step.

\subsubsection{Majorization}

For our multinomial model, we like to minimize the negative log-likelihood, that is
\[
\mathcal{L}(\bs{\theta}) = \sum_i \mathcal{L}_i(\bs{\theta}_i) = -\sum_i \sum_k g_{ik}\log\left[\frac{\exp(\theta_{ik})}{\sum_{k' = 1}^K\exp(\theta_{ik'})} \right],
\]
where \(\bs{\theta}_i = [\theta_{i1}, \ldots, \theta_{iK}]'\) and $g_{ik}$ are the elements of the response profile indicator matrix ($\bm{G}$), rows of which will be indicated by $\bm{g}_i$. For deriving the majorization function we focus on $\mathcal{L}_i(\bs{\theta}_i)$ and afterwards sum the contributions over participants. 

We majorize the negative log-likelihood using a quadratic upper bound \citep[][Section 3.4]{hunter2004tutorial} which is based on the following inequality
\[
\mathcal{L}_i(\bs{\theta}_i) \leq \mathcal{L}_i(\bs{\vartheta}_i) + \frac{\partial \mathcal{L}_i(\bs{\theta}_i)}{\partial \bs{\theta}_i}(\bs{\theta}_i - \bs{\vartheta}_i) + \frac{1}{2}(\bs{\theta}_i - \bs{\vartheta}_i)'\bs{\Omega}(\bs{\theta}_i - \bs{\vartheta}_i), 
\]
for any matrix $\bs{\Omega}$ such that \(\bs{\Omega} - \partial^2 \mathcal{L}_i(\bs{\theta}_i)\) is positive semi
definite. \cite{bohning1988monotonicity}, \cite{bohning1992multinomial}, and \cite{evans2014logistic} showed that the matrix \(\bs{\Omega} = \frac{1}{4}\mathbf{I}\) defines such a matrix. The first derivative of $\mathcal{L}_i(\bs{\theta}_i)$ is
\[
\frac{\partial \mathcal{L}_i(\bs{\theta}_i)}{\partial \bs{\theta}_i} =  - (\mathbf{g}_i - \bs{\pi}_i),
\] 
such that 
\[
\mathcal{L}_i(\bs{\theta}_i) \leq \mathcal{L}_i(\bs{\vartheta}_i) - (\bs{\theta}_i - \bs{\vartheta}_i)'(\mathbf{g}_i - \bs{\pi}_i) + \frac{1}{8}(\bs{\theta}_i - \bs{\vartheta}_i)'(\bs{\theta}_i - \bs{\vartheta}_i). 
\] 

Working out this expression and defining
\(\bm{h}_i = \bs{\vartheta}_i + 4(\mathbf{g}_i - \bs{\pi}_i)\),
we obtain after some simplification
\[
\mathcal{L}_i(\bs{\theta}_i) \leq  \frac{1}{8} \bs{\theta}_i'\bs{\theta}_i -\frac{1}{8}\cdot 2\bs{\theta}_i' \bm{h}_i + \mathrm{constant}
\] 
\citep[see][]{deleeuw2006principal, groenen2003}. Note that \(\bm{h}_i\) is independent of \(\bs{\theta}_i\), so we can add
\(\frac{1}{8}\bm{h}_i'\bm{h}_i\) to obtain 
\[
\mathcal{L}_i(\bs{\theta}_i) \leq  
\frac{1}{8} \bs{\theta}_i'\bs{\theta}_i
-\frac{1}{8}\cdot 2\bs{\theta}_i' \bm{h}_i
+ \frac{1}{8}\bm{h}_i' \bm{h}_i
+ \mathrm{constant},
\] 
which can be rewritten as 
\[
\mathcal{L}_i(\bs{\theta}_i) \leq 
\frac{1}{8} || \bm{h}_i - \bs{\theta}_i||^2 + \mathrm{constant} = \mathcal{M}_i(\bs{\theta}_i| \bm{h}_i) + \mathrm{constant}.
\] 
By using the summation property, we therefore have the following majorization function 
\[
\mathcal{M}(\bs{\theta}|\bs{H}) = \sum_i \mathcal{M}_i(\bs{\theta}_i |\bm{h}_i) = \sum_i \sum_k(h_{ik} - \theta_{ik})^2 = \| \bm{H} - \bm{\Theta} \|^2,
\] 
where $\bm{H} = \{h_{ik}\}$ with \( h_{ik} = \vartheta_{ik} + 4(g_{ik} - \pi_{ik})\), so called \emph{working responses}. The matrix $\bm{\Theta}$ has elements $\theta_{ik}$ and is parametrized as
\[
\bm{\Theta} = \bm{1b}_w'\bm{W}' + \bm{XB}_x\bm{B}'_z\bm{Z}'.
\]
Therefore, the loss function to be minimized in each iteration of the MM algorithm is
\begin{eqnarray}
\| \bm{H} - \bm{1b}_w'\bm{W}' - \bm{XB}_x\bm{B}'_z\bm{Z}' \|^2.
\label{eq:ls_loss}
\end{eqnarray}

\subsubsection{Minimization}\label{sec:updates}

Within each iteration, we have to minimize the least squares loss function (\ref{eq:ls_loss}) which we do in an alternating fashion. To update the parameter vector $\bm{b}_w$, we first define the auxiliary unconstrained vector
\begin{eqnarray*}
\tilde{\bm{m}} = \frac{1}{N} (\bm{H} - \bm{XB}_x\bm{B}'_z\bm{Z}')' \bm{1},
\end{eqnarray*}
from which we find the update as
\begin{eqnarray}
\bm{b}_w^+ = \left(\bm{W}'\bm{W}\right)^{-1}\bm{W}'\tilde{\bm{m}},
\label{eq:updatebm}
\end{eqnarray}
and subsequently compute the update \(\bm{m}^+ = \bm{W}\bm{b}_w^+\).

To find updates for $\bm{B}_x$ and $\bm{B}_z$, we first define
\[
\bm{H}_c = \bm{H} - \bm{1m}', 
\]
to obtain the loss function
\[
\| \bm{H}_c - \bm{XB}_x\bm{B}'_z\bm{Z}' \|^2,
\]
which is the so-called CANDELINC problem \citep{douglas1980candelinc}. Updates of $\bm{B}_x$ and $\bm{B}_z$ can be obtained from a generalized singular value decomposition of the $P \times Q$ matrix
\[
(\bm{X}'\bm{X})^{-1}\bm{X}'\bm{H}_c\bm{Z}(\bm{Z}'\bm{Z})^{-1}
\]
in the metrics $(\bm{X}'\bm{X})$ and $(\bm{Z}'\bm{Z})$ \citep{takane2013constrained}. Therefore, take the SVD of  
\[
(\bm{X}'\bm{X})^{-\frac{1}{2}}\bm{X}'\bm{H}_c\bm{Z}(\bm{Z}'\bm{Z})^{-\frac{1}{2}} = \bm{P}\bm{\Phi}\bm{Q}'
\]
and define the updates
\begin{eqnarray}
\bm{B}_x^+ &=& (\bm{X}'\bm{X})^{-\frac{1}{2}} \bm{P}_S \bm{\Phi}_S, \label{eq:updateBx}\\ 
\bm{B}_z^+ &=& (\bm{Z}'\bm{Z})^{-\frac{1}{2}} \bm{Q}_S, \label{eq:updateBz}
\end{eqnarray}
where $\bm{P}_S$ and $\bm{Q}_S$ are the first $S$ left and right singular vectors, and $\bm{\Phi}_S$ is the $S \times S$ diagonal matrix with the $S$ largest singular values. 

In the coming sections, we will sometimes talk about he maximum dimensionality. This maximum dimensionality is the maximum value of $S$. From the SVD shown above we see that this maximum equals $\min (P, Q)$. Contrary to, for example, principal component analysis, the maximum dimensionality thus not only depends on the data but also on the imposed structure of the model. This structure depends on the two design matrices $\bm{X}$ and $\bm{Z}$. The number of columns of these two matrices determines the maximum dimensionality.

In short, in every iteration of the MM algorithm, we compute the matrix of working responses $\bm{H}$. Then update $\bm{b}_w$ by Equation (\ref{eq:updatebm}) and update $\bm{B}_x$ and $\bm{B}_z$ by Equations (\ref{eq:updateBx}) and (\ref{eq:updateBz}). Having these updates, new working responses are computed. This sequence of steps monotonically minimizes the negative log-likelihood to a global minimum. Note that the algorithm is computationally efficient in the sense that many matrices and their inverses (like, for example, $(\bm{X}'\bm{X})^{-\frac{1}{2}}\bm{X}'$) only need to be computed once, in contrast to Newton or iteratively reweighted least squares algorithms for which the weights and some inverses of matrices have to be computed in every iteration. 

\subsubsection{Alternative minimization procedures}

In the previous subsection, we derived an update for $\bm{B}_x$ and $\bm{B}_z$ simultaneously. To show the versatility of the MM approach, we discuss some alternative updating schemes that also allow for further constraints on the parameters.  One alternative updating strategy can be conceived where each matrix of parameters is updated separately. When updating $\bm{B}_x$, we consider $\bm{B}_z$ fixed so that we may write the majorization function as
\begin{eqnarray*}
\| \bm{H}_c - \bm{XB}_x\bm{V}' \|^2,
\end{eqnarray*}
which is now only a function of $\bm{B}_x$. Minimizing this function is straightforward \citep{penrose1956best, tenberge1993}, and the update is given by
\[
\bm{B}_x^+ = (\bm{X}'\bm{X})^{-1}\bm{X}'\bm{H}_c\bm{V}. 
\]

When updating $\bm{B}_z$, we consider $\bm{B}_x$ fixed so that we may write the majorization function as
\begin{eqnarray*}
\| \bm{H} - \bm{1b}_w'\bm{W} - \bm{UB}'_z\bm{Z}' \|^2,
\end{eqnarray*}
subject to the imposed identification constraint. The update can be found using Kristof's upper bound as has been shown by \cite{tenberge1993}, which is 
\[
\bm{B}_z^+ = (\bm{Z}'\bm{Z})^{-\frac{1}{2}}\bm{P}_S\bm{Q}'_S
\]
where $\bm{P}_S$ and $\bm{Q}_S$ are the left and right singular vectors corresponding to the $S$ largest singular values of the matrix $(\bm{Z}'\bm{Z})^{-\frac{1}{2}}\bm{Z}'\bm{H}'_c\bm{U}$. 

For another alternative updating scheme, we may rewrite the majorization function as
\begin{eqnarray*}
\| \bm{H} - \bm{1b}_w'\bm{W} - \sum_{s=1}^S \bm{Xb}^{(x)}_s(\bm{b}_s^{(z)})' \bm{Z}' \|^2,
\end{eqnarray*}
where $\bm{b}_s^{(x)}$ is the $s$-th column of $\bm{B}_x$ and $\bm{b}_s^{(z)}$ is the $s$-th column of $\bm{B}_z$. This formulation shows that updates can be obtained dimension by dimension. Such dimension wise updating makes it feasible to set some of the elements of  $\bm{b}_s^{(x)}$ or $\bm{b}_s^{(z)}$ to zero, for example, in the case where we have prior information about the structure of the response variables. \cite{derooij2023melodic2}, for example, use this approach where a prior theory states that three response variables pertain to the first dimension and two other response variables pertain to the second dimension. In such a case, researchers might want to fix some of the elements of $\bm{b}_s^{(z)}$ to zero, meaning that the corresponding items do not pertain to dimension $s$. Similarly, some predictors may pertain to certain dimensions for which we might use the same dimension wise updating scheme. Depending on the specific set of constraints we might need to change the identification constraints. 

\subsection{Model selection}

For selecting a model there are four issues:
\begin{itemize}
    \item Selecting the dimensionality/rank $S$;
    \item Selecting the structure on the profile scores ($\bm{v}_k$), that is, determine $\bm{Z}$;
    \item Selecting the linear predictor of the object scores ($\bm{u}_i$), that is, determine $\bm{X}$;
    \item Determining the association structure among the response variables, that is, determine $\bm{W}$.
\end{itemize}

In general, for maximum likelihood methods there are several type of statistics that can be used for inference. The best known statistics are Wald tests, likelihood ratio tests, and information criteria like Akaike's Information Criterion \citep[AIC; ][]{akaike1974new}. 

For our canonical decomposition model, we need to make the following observations. For the computation of the Wald statistics, we need to have an estimate of the standard errors of the parameters. Usually, these are obtained from the Hessian matrix. However, as we use an MM algorithm that does not optimize the log likelihood itself this Hessian is not a by-product of the algorithm. This need not be a problem, confidence intervals for the parameters can be obtained from a bootstrap procedure \citep{efron1979bootstrap, efron1986bootstrap}. The likelihood ratio statistic compares two nested models. The statistic divides the likelihood value obtained under a null hypothesis by the likelihood obtained under an alternative hypothesis. If the model under the null hypothesis is true and certain regularity conditions are satisfied, the likelihood ratio statistic is known to be asymptotically distributed as a chi-square variable with degrees of freedom equal to the difference in the number of parameters under the two hypotheses. For the model proposed in this paper there are two complications: 1) The regularity conditions are not satisfied for selecting the optimal dimensionality, see \cite{takane2003likelihood} and \cite{takane2023likelihood} for a detailed discussion; 2) we generally do not belief a certain model to be true as this involves many assumptions. Suppose that we would like to test whether the first predictor has an effect on the responses and therefore estimate the model with and without this first predictor. Even if the first predictor has no effect on the outcomes, the model can be misspecified in many other aspects. One example of such misspecification is that another predictor has a nonlinear effect on the log odds of one or more response variables.   

Therefore, we will use the AIC for selection of an optimal model. AIC is based on the entropic or information-theoretic interpretation of the maximum likelihood method as well as the minimization of the Kullback-Leibler information quantity \citep{akaike1974new, burnham2004multimodel, anderson2007model}. The AIC for any model can be defined as
\[
    \mathrm{AIC} =  2 \hat{\mathcal{L}}(\bm{\theta}) + 2\mathrm{npar},
\]
where npar denotes the number of parameters of the model. The first term $2\hat{\mathcal{L}}(\bm{\theta})$ in AIC is twice the negative log likelihood (usually called the deviance) and it acts as a measure of lack of fit to the data, consequently smaller values will be preferred. The second term, $2$npar, acts as a penalty term which penalizes complex models having many parameters. The aim is to reach a balance between the lack of fit and the model complexity: models with smaller AIC values will indicate a better balance. The optimal model choice minimizes the AIC. For the AIC, the number of parameters is needed, we will use $\mathrm{npar} = T + S(P + Q - S)$.  

As there are many choices to make, we follow the suggestion of \cite{yu2013model} to use a step-wise approach. In the step-wise approach, we  start with full matrices $\bm{Z}$,  $\bm{X}$, and $\bm{W}$. That is, define $\bm{Z}$ and $\bm{W}$ to include all main effects and associations among the responses, and defined $\bm{X}$ to have all possible predictors and interactions of interest. The first selection step is determining the dimensionality $S$. Second, using the just selected $S$, find the optimal profile structure, that is, find the optimal design matrix $\bm{Z}$, by leaving out columns pertaining to higher order associations. Third, with the optimal $S$ and profile structure, find which predictor variables contribute to the prediction of the responses, that is, find the optimal $\bm{X}$ by leaving out columns of $\bm{X}$. Finally, determine which associations among the response variables are necessary, that is, find the optimal $\bm{W}$. We will illustrate this model selection procedure in Section \ref{sec:nesda}.

\section{Details for profiles of binary response variables}\label{sec:dichotomous}

The canonical decomposition model, as laid-out in the previous section, models the probability of an outcome of the categorical response variable $\mathcal{G}$. Let us denote the conditional probability of a response variable given the participant predictors as $\pi_{k}(\bm{x}_i) = P(\mathcal{G}_i = k|\bm{x}_i)$. In our model this probability equals 
\begin{eqnarray*}
\pi_{k}(\bm{x}_i) = \frac{\exp(m_k + \bm{x}_i'\bm{a}_k)}{\sum_{k' = 1}^K \exp(m_{k'} + \bm{x}_i'\bm{a}_{k'})}, 
\end{eqnarray*}
where $m_k = \bm{w}_k'\bm{b}_w$ and $\bm{a}_k = \bm{B}_x\bm{B}'_z\bm{z}_k$. When the response categories represent profiles of response variables, using this equation, we can directly derive the probability of a response profile given the predictor variables of a participant. This derivation follows the rules of multinomial logistic regression models. A researcher, however, might be interested not only in the probability of the profile, but also in the probability or the log odds of one of the underlying binary response variables (as in a binary logistic regression). Furthermore, a researcher might be interested in the effect of a predictor on the association between two binary response variables. In this section, we discuss how to obtain such effects from our model and derive interpretational rules for these log odds and  log odds ratios. First, we discuss the design matrices $\bm{W}$ and $\bm{Z}$ in more detail. Afterwards, we derive the rules of interpretation. 

\subsection{Definition of design matrices}

For multivariate binary data, the $K$ response categories of \resp are combinations of the $R$ binary variables. Having, say, 3 response variables with answer categories correct (C) and wrong (W), one of the possible profiles is (C, C, W), that is, the first two correct and the third wrong. Coding the binary response variables with \(y_r = -\frac{1}{2}\) for one of the classes (say, wrong) and \(y =  +\frac{1}{2}\) for the other (say, correct), we can define the design matrices $\bm{Z}$ and/or $\bm{W}$. Note that $\bm{Z}$ is used to represent relationships between the predictor variables and the response variables, whereas $\bm{W}$ is used to represent relationships among different response variables. 

Suppose, we decide to have a main effect of response variable $r$ in the design matrix $\bm{Z}$. This means that the predictors have an effect on this response variable. In that case, we define a column of $\bm{Z}$ with corresponding elements \(y_r\). Furthermore, suppose we decide to have an association between response variable $r$ and $r'$ in the design matrix $\bm{Z}$. A column of this matrix is than defined by the product \(y_r y_{r'}\). Including this column means that the predictors have an effect on the association between the two response variables. To obtain a hierarchical model, we always include the underlying main effects in the design matrix when an association is included. 

As an example, consider $R = 3$ underlying binary response variables such that $K = 2^3 = 8$. A possible design matrix having main effects, two-variable associations, and a three-variable association is given by
\begin{eqnarray}
\bm{Z} = \begin{bmatrix*}[r]
-\frac{1}{2} & -\frac{1}{2} & -\frac{1}{2} &  \frac{1}{4} &  \frac{1}{4} &  \frac{1}{4} & -\frac{1}{8} \\ 
-\frac{1}{2} & -\frac{1}{2} &  \frac{1}{2} &  \frac{1}{4} & -\frac{1}{4} & -\frac{1}{4} &  \frac{1}{8} \\ 
-\frac{1}{2} &  \frac{1}{2} & -\frac{1}{2} & -\frac{1}{4} &  \frac{1}{4} & -\frac{1}{4} &  \frac{1}{8} \\ 
-\frac{1}{2} &  \frac{1}{2} &  \frac{1}{2} & -\frac{1}{4} & -\frac{1}{4} &  \frac{1}{4} & -\frac{1}{8} \\ 
\frac{1}{2} & -\frac{1}{2} & -\frac{1}{2} & -\frac{1}{4} & -\frac{1}{4} &  \frac{1}{4} &  \frac{1}{8} \\ 
\frac{1}{2} & -\frac{1}{2} &  \frac{1}{2} & -\frac{1}{4} &  \frac{1}{4} & -\frac{1}{4} & -\frac{1}{8} \\ 
\frac{1}{2} &  \frac{1}{2} & -\frac{1}{2} &  \frac{1}{4} & -\frac{1}{4} & -\frac{1}{4} & -\frac{1}{8} \\ 
\frac{1}{2} &  \frac{1}{2} &  \frac{1}{2} &  \frac{1}{4} &  \frac{1}{4} &  \frac{1}{4} &  \frac{1}{8}
\end{bmatrix*}.
\label{eq:Zdesign}
\end{eqnarray}
The first three columns represent each of the three underlying response variables, columns 4 till 6 represent two-variable associations ($y_{r}y_{r'}$), and the last column defines a three variable association ($y_{r}y_{r'}y_{r''}$) (where $r'$ and $r''$ are $1,\ldots, R$ but not equal to each other). This design matrix is a saturated design matrix for three binary response variables. In the model selection procedure, a researcher can select a subset of the columns of this design matrix. 

With the design matrix $\bm{Z}$ a structure is imposed on the profile scores  ($\bm{v}_k$). By imposing the structure, we define specific relationships between on the one hand the predictor variables ($\bm{x}_i$) and on the other hand the responses or associations among the responses. We could, for example, impose a \emph{main effect structure} on the profiles scores. In that case, we define the design matrix $\bm{Z}$ by only the first three columns of the table above, such that $z_{k1} = y_1$, $z_{k2} = y_2$, and $z_{k3} = y_3$. The profile scores for the $k$-th profile become
\begin{equation}
v_{ks} = \sum_r z_{kr} b_{rs}^{(z)},
\label{eq:main}
\end{equation}
where $b_{rs}^{(z)}$ is an element of the matrix $\bm{B}_z$. Alternatively, a \emph{main effect and two-way association structure} can be imposed, where we use the first six columns of design matrix (\ref{eq:Zdesign}). Then, $z_{k1} = y_1$, $z_{k2} = y_2$, $z_{k3} = y_3$, $z_{k4} = y_1y_2$, $z_{k5} = y_1y_3$, and $z_{k6} = y_2y_3$, so that
\begin{equation}
v_{ks} = \sum_r z_{kr} b_{rs}^{(z)} + \sum_{r, r' > r} z_{kr} z_{kr'} b_{rr's}^{(z)} = \sum_r z_{kr} \left(b_{rs}^{(z)} + \sum_{r' > r} z_{kr'} b_{rr's}^{(z)}\right),
\label{eq:mainandassoc}
\end{equation}
where $rr' = 4$ when $r = 1$ and $r' = 2$, 
$rr' = 5$ when $r = 1$ and $r' = 3$, and 
$rr' = 6$ when $r = 2$ and $r' = 3$. 
There are also other possibilities, such as all main effects and a few two-way associations or extending to higher order associations (i.e., also including the last column). The specification of $\bm{Z}$ depends on the research question. Generally, the $K \times Q$ design matrix $\bm{Z}$ includes as columns the relevant effects, such that we can write $\bm{V} = \bm{Z}\bm{B}_z$, where the weights $\bm{B}_z$ are parameters to be estimated. 

The intercepts ($m_k$, $k = 1,\ldots,K$) model the associations among the response variables. Without constraints on these intercepts, a saturated model is assumed for this part of the data. However, there might only be pairwise associations among the response variables. A $K \times T$ design matrix $\bm{W}$ can be used that imposes a structure on the relationships among the response variables. The columns of $\bm{W}$ are defined similarly to the design matrix $\bm{Z}$, see (\ref{eq:Zdesign}). If only the first three columns are used, we assume the responses to be conditionally independent. If the first six columns are used, we assume there are pairwise associations among the responses but no three-variable association. To obtain a hierarchical overall model, the columns of $\bm{Z}$ (pertaining to the associations between predictor and response variables) should be a subset of the columns of $\bm{W}$ (pertaining to the associations among responses).

\subsection{Modelling the log odds}\label{sec:logodds}

Suppose we like to know the effect of the $p$-th predictor on the $r$-th response variable. In binary logistic regression, such a effect is defined by the change in estimated log odds corresponding to a unit change in the predictor. We will develop a similar rule for our model. 

Therefore, let us consider the following two profiles
\( k = (\bullet, \  \mathrm{C}, \ \bullet) \) and \( l = (\bullet, \ \mathrm{W}, \ \bullet) \), that only differ in the $r$-th (= 2) position, that is, the dots indicate that the corresponding responses in the profile can take any value as long as they are the same in the two profiles. The log odds for this $r$-th variable is given by $\log(\frac{\pi_{ik}}{\pi_{il}})$. This log odds depends on the structure we impose on the profile scores but also on the value of the predictor variable. To make this latter dependency clearer, we will denote $\pi_{ik} = \pi_{k}(\bm{x}_i)$, that is, the probability for profile $k$ depends on the predictor variables $\bm{x}_i$. 

In our model, the following expression pertains to this log odds 
\begin{equation}
\lambda_r(\bm{x}_i) = \log \left( \frac{\pi_k(\bm{x}_i)}{\pi_l(\bm{x}_i)} \right) = \log(\pi_k(\bm{x}_i)) - \log(\pi_l(\bm{x}_i)) = m_k - m_l + \bm{x}'_i \bm{B}_x(\bm{v}_k - \bm{v}_l).
\label{eq:logodds}
\end{equation}

From this expression, we see that when $\bm{x}_i = \bm{0}$, the log odds only depend on the estimated intercepts $m_k$ and $m_l$. These intercepts are parameterized using the design matrix $\bm{W}$. This log odds can be further worked out, but the expression depends on the structural form imposed on the profile scores through the design matrix $\bm{Z}$. Using Equations (\ref{eq:main}) and (\ref{eq:mainandassoc}) the following expressions are obtained:
\begin{itemize}
\item With a \emph{main effect structure} on the profile scores, we have for the last term in (\ref{eq:logodds})
\[
(\bm{v}_k - \bm{v}_l) = \bm{B}'_z (\bm{z}_k - \bm{z}_l) = (z_{kr} - z_{lr}) \sum_s b^{(z)}_{rs}, 
\]
because only the $z$-values of the $r$-th response variable differ between the two profiles (i.e., the others are equal, see equation \ref{eq:main}). Furthermore, \( z_{kr} - z_{lr}  = 1\). Therefore, when $x_p$ increases with a unit, the log odds of response variable $r$, that is, \( \lambda_r(\bm{x}_i) \), goes up by $\sum_s b_{ps}^{(x)} b_{rs}^{(z)}$. 
\item With a \emph{main and association effect structure} on the profile scores it follows from Equation (\ref{eq:mainandassoc}) that the last term in (\ref{eq:logodds}) becomes
\[
(\bm{v}_k - \bm{v}_l) = (z_{kr} - z_{lr}) \sum_s \left(b_{rs}^{(z)} + \sum_{r' > r} z_{kr'} b_{rr's}^{(z)}\right),   
\]
which does not only depend on the main effect parameters anymore. Therefore, with this structure, the change in log odds for response variable $r$ corresponding to a unit increase in a predictor has no simple expression anymore. Consequently, model interpretation should focus on the highest order effects, similar to loglinear models \citep[][Section 8.2.3]{agresti2002categorical}. 
\end{itemize}

\subsection{Modelling the log odds ratio}\label{sec:logoddsratio}

Besides inspecting the effect of a predictor on a single response, we might also be interested in the effect of the $p$-th predictor on the association between response variables $r$ and $r'$. The association between two categorical response variables is usually defined in terms of the (log) odds ratio. The effect of a predictor variable is then defined as the change in the log odds ratio corresponding to a unit change in the predictor. We will develop such a rule for our canonical decomposition model. 

Therefore, let us consider the following four profiles
\( k = (\bullet, \ \mathrm{C}, \mathrm{C})\), \( l = (\bullet, \ \mathrm{W}, \mathrm{C}) \), 
\( n = (\bullet, \ \mathrm{C}, \mathrm{W}) \), and \( o = (\bullet, \ \mathrm{W}, \mathrm{W}) \), 
that only differ in the $r$ (= 2) and $r'$-th (= 3) position, that is, the dots indicate that the corresponding responses in the profile can take any value as long as they are the same in the four profiles. The log odds ratio for the association between the $r$-th and $r'$-th variable is given by $\log(\frac{\pi_{ik}\pi_{io}}{\pi_{il}\pi_{in}})$. Again, this log odds ratio depends on the value of the predictor variable and the structure we impose on the profile scores. 

In our canonical decomposition model the following expression pertains to this log odds ratio 
\begin{eqnarray}
\kappa_{r, r'}(\bm{x}_i) &=& \log \left( \frac{\pi_k(\bm{x}_i) \pi_o(\bm{x}_i)}{\pi_l(\bm{x}_i) \pi_n(\bm{x}_i)} \right) = \log(\pi_k(\bm{x}_i)) + \log(\pi_o(\bm{x}_i)) -  \log(\pi_l(\bm{x}_i)) -  \log(\pi_n(\bm{x}_i)) \\
&=& m_k + m_o - m_l - m_n + \bm{x}'_i \bm{B}_x(\bm{v}_k + \bm{v}_o - \bm{v}_l - \bm{v}_n).
\label{eq:logoddsratio}
\end{eqnarray}
Like for the log odds, we see that when $\bm{x}_i = \bm{0}$, the log odds ratio only depend on the estimated intercepts, that are parameterized using the design matrix $\bm{W}$. This expression for the log odds ratio depends on the imposed structure on the profile scores. Using Equations (\ref{eq:main}) and (\ref{eq:mainandassoc}) we get
\begin{itemize}
    \item with a \emph{main effect structure} on the profile scores, all $\kappa_{r, r'}(\bm{x}_i) = m_k + m_o - m_l - m_n$, that is, the log odds ratio does not depend on the predictors.
    \item with a \emph{main and association effect structure} we obtain
    \[
    (\bm{v}_k + \bm{v}_o - \bm{v}_l - \bm{v}_n) = \bm{B}_z' (\bm{z}_{k} + \bm{z}_{o} - \bm{z}_{l} - \bm{z}_{n}) = (z_{kr}z_{kr'} + z_{or}z_{or'} - z_{lr}z_{lr'} - z_{nr}z_{nr'})\sum_s b_{rr's}^{(z)} = \sum_s b_{rr's}^{(z)},
    \]
    so that, when $x_p$ increases with a unit, the log odds ratio \( \kappa_{r,r'}(\bm{x}_i) \) goes up by $\sum_s b_{ps}^{(x)} b_{rr's}^{(z)}$. 
\end{itemize} 

\section{Closely related models}\label{sec:relations}

In this section, we will discuss relationships of our approach with other analysis methods. We will point out relationships with loglinear models, multinomial logistic regression, reduced rank multinomial logistic regression, and double constrained correspondence analysis. 

\subsection{Loglinear analysis}\label{sec:complla}

When all predictor variables and response variables are categorical, we can form a contingency table of the $P + R$ variables. Loglinear analysis can be used to investigate the conditional associations among the variables. No distinction is made between predictor and response variables, although in applying the method a researcher might make such a distinction. In a loglinear analysis, a researcher can choose to include any association (pairwise or higher order) between variables, and can thus see whether the log odds of one variable depends on another variable, or whether the log odds ratio between two variables depends on a third one (i.e., compare Section \ref{sec:dichotomous}). Usually loglinear models are constrained to be hierarchical, in the sense that when an association is included in the model all lower order associations between and main effects of the variables are also included

For purely categorical predictors and responses, our model in maximum dimensionality (i.e., $\min(P, Q)$) is equivalent to certain loglinear models. This equivalence is in terms of the same modelled log odds or log odds ratios. Furthermore, differences between deviances of our models correspond to differences in deviance of equivalent loglinear models. We will illustrate using an empirical data set in the next Section. 

Let us discuss some similarities and differences between loglinear models and our canonical decomposition model. Similar to logistic regression models, in our model, we do not make any assumption about the relationships among the predictor variables. When comparing the two approaches, in the loglinear model we need all associations among the predictor variables. Second, in loglinear analysis any association term among response variables can be in or excluded. In our approach, we define them using the design matrix $\bm{W}$. Finally, the core of our model is in the relationships between predictors and responses, and is captured by the the design matrices $\bm{X}$ and $\bm{Z}$ and corresponding coefficients $\bm{B}_x$ and $\bm{B}_z$. 

A critical difference between loglinear analysis and our approach is that in our model we do not have the possibility to specifically define which predictor variable has an influence on which outcome variable, whereas in loglinear models such a specification is possible. When a predictor variable is included in the model through the design matrix $\bm{X}$, it will have an effect on all responses and associations as defined in $\bm{Z}$. Therefore, not all loglinear models have an equivalent representation in our canonical decomposition model. Specifically, if a main effect of the first predictor is included defined in $\bm{X}$ and we define the matrix $\bm{Z}$ to include all main effects and the association among response variables 1 and 2, the first predictor has an effect on each response variable and an effect on the association between the first two response variables, as pointed out in Section \ref{sec:logoddsratio}. 

When we compare the (difference in) deviances, we need to take into account that the loglinear model operates with so-called grouped data, whereas our model works with ungrouped data \cite[see][Section 5.2.3]{agresti2013categorical}. The deviance for those two forms of data is different, even when they represent the same information. For grouped data, the deviance can be used as a model fit statistic, which is not the case for ungrouped data. When comparing the results of loglinear models with the results of our model, the deviance values cannot be simply compared. However, differences in deviance between two models will be the same for loglinear models and our canonical decomposition model.

Whereas loglinear models are only feasible with categorical data, our canonical decomposition model can also be used when we have numerical predictors for the participants or numerical characteristics of the profiles. Furthermore, in our model we have a probability for every profile of the response variables that cannot be obtained from a loglinear analysis. 

\subsection{Multinomial logistic regression}

In the standard multinomial regression model $\theta_{ik}$ is parameterized as $\theta_{ik} = m_k + \bm{x}'_i\bm{a}_k$, where $\bm{a}_k$ is a vector with regression weights for response profile $k$. Usually, one of the elements $m_k$ and corresponding vector $\bm{a}_k$ are fixed to zero for identification. 

Multinomial logistic regression is a special case of our model. To see that more clearly, the canonical parameters of multinomial logistic regression can be written as
\[
\bm{\Theta} = \bm{1m}' + \bm{XA},
\]
whereas for our canonical decomposition it is
\[
\bm{\Theta} = \bm{1b}_w'\bm{W}' + \bm{XB}_x\bm{B}'_z\bm{Z}'.
\]
If we define the design matrices $\bm{Z}$ and $\bm{W}$ as ``saturated matrices'' and choosing $S$ to be equal to $\min(P, K-1)$, the canonical parameters ($\theta_{ik}$) and therefore the probabilities $\pi_{ik}$ will be exactly equal between the two modelling approaches. 

A multinomial logistic regression might, however, not always be feasible for the data that we consider. That is, $K$ is an exponential function of the number of response variables and grows quickly, that is, with 8 response variables $K$ is already 256 and with 11 response variables $K$ equals 2048. When $K$ is large, many parameters need to be estimated for this model (i.e., $(P +1) \times (K - 1)$), that is, there are $K - 1$ vectors $\bm{a}_k$ of length $P$. This might become an enormous amount of parameters to be estimated and interpreted. 

\subsection{Multinomial reduced rank logistic regression}

\cite{anderson1984regression} and \cite{yee2015} defined multinomial reduced rank regression models. The stereotype model \citep{anderson1984regression} is a special case with rank equal to one. In this multinomial reduced rank regression model, the canonical parameters are 
\[
\bm{\Theta} = \bm{1m}' + \bm{XB}_x\bm{V}',
\]
where $\bm{V}$ is a $K \times S$ matrix with parameters. 
This representation shows that our model is closely related to this model. In the canonical decomposition model, further restrictions on the matrix with profile scores $\bm{V}$ are imposed using the matrix $\bm{Z}$. If $\bm{W}$ and $\bm{Z}$ are saturated design matrices, our canonical decomposition becomes equivalent to the multinomial reduced rank regression model, that is, the estimated canonical parameters and probabilities will be equal. The number of parameters in a multinomial reduced rank regression model is $K - 1 + (P + K - S)S$, whereas for our canonical decomposition model it is $T + (P + Q - S)S$. Because usually we will have $T \leq K - 1$ and $Q < K$, a substantial reduction of parameters can be achieved. 

\subsection{Double constrained correspondence analysis}

Another related data analysis technique is double constrained correspondence analysis \citep{terbraak2018algorithms}. Canonical correspondence analysis \cite{terbraak1986canonical, terbraak1987analysis} is an extension of classical correspondence analysis \citep{greenacre1984} where the row scores are linear combinations of external variables. Double constrained correspondence analysis can be considered a similar extension of canonical correspondence analysis, where not only the row scores are 
linear combinations of external variables, but also the column scores are linear combinations of a set of other external variables. 

Similarly, we linearly constraint the object scores and profile scores with external information. Whereas, double constrained correspondence analysis is a least squares technique, we developed a maximum likelihood method. Furthermore, double constrained correspondence analysis is often applied to matrices that contain counts, for example, the number of a species that are observed in a specific environment. As far as we know, double constrained correspondence analysis has not been applied to a data matrix having only a single one per row, as our main matrix $\bm{G}$ has. Our modelling of log odds and log odds ratios in Sections \ref{sec:logodds} and \ref{sec:logoddsratio} are similar to what is called the \emph{fourth corner correlation} in double constrained correspondence analysis \citep{terbraak2018algorithms}, the  correlation that is maximized by that technique. 

\section{Empirical examples}\label{sec:applications}

In this section, we will show two examples. The first analysis has three dichotomous response variables and two dichotomous predictor variables. We use these data to illustrate the relationships between our canonical decomposition model and loglinear analysis as described in Section \ref{sec:relations}. Equivalence relationships between the two analysis approaches will be shown in detail. Not all loglinear models, however, can be cast as a canonical decomposition model. We will discuss which loglinear models have and which do not have an equivalent canonical decomposition form. As for purely categorical variables, the loglinear approach is more general we conclude that in this case (i.e., when all variables are categorical) a researcher better uses loglinear models for the statistical analysis. 

However, when there are also some numerical predictor variables, loglinear models cannot be applied anymore. In that case, the canonical decomposition models as introduced in this paper will become useful. The second example 
has five dichotomous response variables and eight predictors. From the eight predictor variables, seven are continuous and one is dichotomous. This is a typical example, for which the multinomial canonical decomposition model was developed. We illustrate the method in detail, starting with model selection and ending with model interpretation.

\subsection{Loglinear \emph{versus} Canonical decomposition model}\label{sec:ACMdata}

Agresti (2013, Table 10.1) shows data of 2276 senior high school students and their use of alcohol (A), cigarettes (C), and marijuana (M), all binary yes or no. Also available are two predictor variables, race (R; white/other) and gender (G; female/male). This data set with three responses and two predictor variables has been analyzed by Agresti using loglinear models. In this Section, we re-analyze the data using our canonical decomposition models. The goal of this comparison is to show relationships between the two approaches, we refrain from substantial interpretations. Table \ref{tab:comparison} shows the main results. 

\begin{table}[t]
\centering
\begin{tabular}{ll|rr | rr}
  & &  \multicolumn{2}{c|}{LLA} & \multicolumn{2}{c}{MCDM} \\ 
  & Model & $G^2$ & $df$ & Deviance & npar \\ \hline  \hline
1 & Mutual independence + GR        & 1325.14   & 25 & 7900.19  & 3 \\
2 & Homogeneous Association         & 15.34     & 16 & 6590.38  & 12\\
3 & All three-factor terms          & 5.27      & 6  & -        & - \\
4a & (2) - AC                        & 201.20    & 17 & 6776.24  & 11\\
4b &(2) - AM                        & 106.96    & 17 & 6681.99  & 11\\
4c &(2) - CM                        & 513.47    & 17 & 7088.51  & 11\\ 
4d &(2) - AG                        & 18.72     & 17 & -        & - \\
4e &(2) - AR                        & 20.32     & 17 & -        & - \\
4f &(2) - CG                        & 16.32     & 17 & -        & - \\ 
4g &(2) - CR                        & 15.78     & 17 & -        & - \\
4h &(2) - GM                        & 25.16     & 17 & -        & - \\
4i & (2) - GR                        & 18.93     & 17 & -        & - \\
5 & (AC, AM, CM, AG, AR, GM, MR, GR)& 16.74     & 18 & 6591.77  & 10\\ 
6 & (AC, AM, CM, AG, AR, GM, GR)    & 19.91     & 19 & -        & - \\
7 & (AC, AM, CM, AG, AR, GR)        & 28.81     & 20 & 6603.84  & 8 \\ \hline \hline
\end{tabular}
\caption{Results from loglinear analyses (LLA) and the multinomial canonical decomposition model (MCDM) for the alcohol (A), cigarettes (C), and marijuana data (M). Predictors are gender (G) and race (R). $G^2$ is the likelihood ratio chi-square statistic, $df$ are the degrees of freedom for this statistic. The deviance is twice our loss function ($2\mathcal{L}(\bs{\theta})$) and npar denotes the number of parameters.} 
\label{tab:comparison}
\end{table}

The left column of Table \ref{tab:comparison} shows all loglinear models as fitted by Agresti. The middle block shows likelihood ratio chi-square statistics $G^2$ for the different models, that is defined as 
\[
G^2 = \sum_{\mathrm{cells}} n_{\mathrm{cell}} \log \frac{n_{\mathrm{cell}}}{\mu_{\mathrm{cell}}},
\]
where $n_{\mathrm{cell}}$ is the observed count in a cell of the contingency table and $\mu_{\mathrm{cell}}$ the corresponding expected value given the loglinear model. Accompanying the $G^2$ statistics are the associated degrees of freedom ($df$) as reported in Table 10.2 of Agresti. Note that the contingency table has 32 cells, the number of parameters of the models is therefore, $32 - df$. The last two columns show the deviance ($2\mathcal{L}(\hat{\bs{\theta}})$) and number of parameters for corresponding canonical decomposition models. 

The equivalent of loglinear model 1 (i.e., mutual independence + GR) can be obtained by setting $S = 0$ and defining $\bm{W}$ to only include the main effects (so, the number of columns of $\bm{W}$ equals $T = 3$). The number of parameters is 3, the length of $\bm{b}_w$. Note that the four parameters representing the association structure among the predictors in the loglinear model are not counted as such in our canonical decomposition model. In the loglinear model there are 7 parameters (32 - 25), that is, the four parameters for the relationships among the predictors (one intercept, two main effects, and 1 association term) and the three main effect parameters for the responses. 

Loglinear model 2, the model with all pairwise associations among the five variables can be obtained as a canonical decomposition model. The definition of $\bm{W}$ should include all three main effects and all three pairwise associations ($T = 6$), but exclude the three-variable association. The design matrix $\bm{Z}$ should include only the main effects ($Q = 3$) of the three responses and $\bm{X}$ should have only the two main effects of race and gender (but not an interaction between them, such that $P = 2$). With this specification, the maximum dimensionality is 2 ($\min(P,Q)$) which we will use ($S = 2$, i.e., the rank is not reduced). The number of fitted parameters in our model is twelve, nine more than in Model 1. The difference between the degrees of freedom of the first two loglinear models is 9, the same difference as between the number of parameters of the first two canonical decomposition models. Although the $G^2$ and deviance values are quite different between the loglinear models and our models (i.e., corresponding to grouped and ungrouped data), it can be verified that 1325.14 - 15.34 = 1309.8 the same difference as 7900.19 - 5590.38 = 1309.8. 

Loglinear model 3 has all three-factor associations. This loglinear model has no equivalent representation using our canonical decomposition model. In the canonical decomposition model, we define the predictor response relationships with the two design matrices $\bm{X}$ and $\bm{Z}$. If a main effect or an interaction effect is defined in $\bm{X}$ it affects all responses or associations as defined in $\bm{Z}$. Similarly, if a response or an association between responses is present in the design matrix $\bm{Z}$, it is affected by all predictor effects defined in $\bm{X}$. Specifically, for loglinear model 3, consider the three-factor associations GRA and GAC. To include GRA in our canonical decomposition, we should have the main effects of and the interaction between gender and race in the design matrix $\bm{X}$ and the main effect of alcohol in the design matrix $\bm{Z}$. To include GAC in our canonical decomposition, we should have the main effect of gender in the design matrix $\bm{X}$ and the main effects of alcohol and cigarettes plus their interaction in the design matrix $\bm{Z}$. Combining these two effects in one model, we obtain a design matrix $\bm{X}$ including the interaction between the two predictors and a design matrix $\bm{Z}$ including the interaction between the two responses. However, in that case, the four-factor interaction GRAC is also included in the model. The all three-factor association model therefore has no equivalent representation in our canonical decomposition family. 

In loglinear models 4a, 4b, and 4c, we simplify the homogeneous association model (model 2) by removing pairwise associations among the responses. In the canonical decomposition model this is achieved by deleting a column from the design matrix $\bm{W}$. It can be verified that the differences in $G^2$ statistics of models 4a, 4b and 4c to the $G^2$ statistic of model 2 are equal to the differences in deviances of these models.

Loglinear models 4d till 4i have no equivalent representation in our family. Compared to model 2, these exclude a pairwise association among a predictor and a response. In the canonical decomposition model, an exclusion of a predictor variable (from $\bm{X}$) implies the effect to all responses is removed from the model. Similarly, a removal of a response variable effect (from $\bm{Z}$) in the canonical decomposition model implies the removal of all predictor effects on this response. 

Agresti removes the first the CR association and then the CG association (model 5). In model 5, there are no effects of the predictors (R and G) on cigarette use anymore. This corresponds to a design matrix $\bm{Z}$ that does not have a main effect of cigarettes anymore, so only main effects of alcohol and marijuana (i.e., $Q = 2$). This model has an equivalent representation in the canonical decomposition model. Differences in $G^2$ statistics and differences in deviance between model 5 and other models are again equal. 

Model 6 deletes the effect of race on marijuana use, which has no canonical decomposition form. Model 7 also deletes the effect of gender on marijuana use. In the canonical decomposition model this can be defined by also removing the column pertaining to marijunana use from the design matrix $\bm{Z}$. As $Z$ now only has one column left (i.e., $Q = 1$), the maximum dimensionality becomes 1. Differences in $G^2$ statistics and differences in deviance between models 5 and 7 are again equal.   

As can be seen in Table \ref{tab:comparison}, several loglinear models have no equivalent canonical decomposition representation (those indicated by ``-''). So, the loglinear framework is more flexible. For all models that have equivalent representations, differences in $G^2$ for the loglinear models are equal to differences in deviances for our models. Differences in degrees of freedom match differences in number of parameters. 

Also estimated effects are the same. Consider model 5, for example, the estimated effect of race of alcohol use is 0.45 in both models, the estimated association among alcohol and cigarette use equals 2.05 in both models. We will not further delve into the interpretation of the effects. A detailed interpretation of the results of loglinear analysis can be found in \cite{agresti2013categorical}.

\subsection{Depressive and anxiety disorders}\label{sec:nesda}

In this section, we will analyse a data set from the Netherlands Study on Depression and Anxiety \citep[NESDA;][]{penninx2008netherlands}, where we have for 786 participants data available on five psychiatric diagnosis (dysthymia (D), major depressive disorder (M), generalized anxiety disorder (G), social phobia (S), and panic disorder (P)). Participants are either having the diagnosis or not. Psychiatric disorders commonly co-occur, a phenomenon called comorbidity. In this particular data set, 272 participants have one disorder, 235 have two disorder, 147 have three, 96 have four, and 36 participants have all five disorders. 

Research focuses on the link between personality characteristics and these psychiatric disorders \citep{spinhoven2009role}. Measurements of the big-five personality characteristics neuroticism (N), extraversion (E), openness to experience (O), agreeableness (A), and conscientiousness (C) are available and will be used as predictors. Besides those five personality characteristics, gender, age and education are available as control variables, which will be included in all models. The variable gender is dichotomous with code 1 for females, the other predictor variables are centered and scaled to have standard deviation equal to one. 


\begin{table}[t]
    \centering
    \begin{tabular}{llllr}
     $S$  & $Z$         & $X$     & $W$ &   AIC \\ \hline \hline
     5  & \bf{3}            & N + E + O + A + C  & \bf{5} &  4449.2\\
     4  & \bf{3}            & N + E + O + A + C  & \bf{5} &  4421.8\\
     3  & \bf{3}            & N + E + O + A + C  & \bf{5} &  4404.6\\
     2  & \bf{3}            & N + E + O + A + C  & \bf{5} &  \emph{4392.2}\\
     1  & \bf{3}            & N + E + O + A + C  & \bf{5} &  4392.5\\ \hline
     2  & \bf{2}            & N + E + O + A + C  & \bf{5} &  4379.0\\
     2  & \bf{1}            & N + E + O + A + C  & \bf{5} &  \emph{4363.3}\\ 
     2  & {\bf 1} + D:M    & N + E + O + A + C  & \bf{5} &  4366.7\\ 
     2  & {\bf 1} + D:G    & N + E + O + A + C  & \bf{5} &  \emph{4361.7}\\ 
     2  & {\bf 1} + D:S    & N + E + O + A + C  & \bf{5} &  4366.9\\
     2  & {\bf 1} + D:P    & N + E + O + A + C  & \bf{5} &  4364.7\\
     2  & {\bf 1} + M:G    & N + E + O + A + C  & \bf{5} &  4362.5\\
     2  & {\bf 1} + M:S    & N + E + O + A + C  & \bf{5} &  4365.2\\
     2  & {\bf 1} + M:P    & N + E + O + A + C  & \bf{5} &  4367.1\\
     2  & {\bf 1} + G:S    & N + E + O + A + C  & \bf{5} &  4366.4\\
     2  & {\bf 1} + G:P    & N + E + O + A + C  & \bf{5} &  4367.1\\
     2  & {\bf 1} + S:P    & N + E + O + A + C  & \bf{5} &  4365.2\\ \hline
     2  & {\bf 1} + D:G    & E + O + A + C  & \bf{5} &  4430.2\\ 
     2  & {\bf 1} + D:G    & N + O + A + C  & \bf{5} &  4378.7\\ 
     2  & {\bf 1} + D:G    & N + E + A + C  & \bf{5} &  \emph{4358.0}\\ 
     2  & {\bf 1} + D:G    & N + E + O + C  & \bf{5} &  4359.2\\ 
     2  & {\bf 1} + D:G    & N + E + O + A  & \bf{5} &  4363.9\\ 
     2  & {\bf 1} + D:G    & N + E + C  & \bf{5} &  \emph{4355.7}\\ \hline
     2  & {\bf 1} + D:G    & N + E + C  & \bf{4} &  4353.7\\ 
     2  & {\bf 1} + D:G    & N + E + C  & \bf{3} &  4347.0\\ 
     2  & {\bf 1} + D:G    & N + E + C  & \bf{2} &  \bf{4336.0}\\ \hline \hline
\end{tabular}
    \vspace{0.7cm}
    \caption{Model Selection for Nesda data. $S$ indicates the dimensionality or rank. The column labeled $Z$ refers to the structure on the profiles scores, where a 3 indicates all main effects plus all two- and three-way associations, 2 indicates all main effects and all two-way associations, and 1 indicates all main effects. 1 + D:M indicates all main effects plus the two-way association between D and M (Dysthymia and Major depressive disorder). Under $X$ it is indicated which predictor variables are in the model. The column labeled $W$ refers to the structure on the associations among the response variables, where the same notation is used as for the profile scores (column $Z$). The final column gives the AIC of the model.}
    \label{tab:nesda}
\end{table}

The results of the model selection procedure are shown in Table \ref{tab:nesda}. The different steps are indicated by the horizontal lines in the table. We will discuss model selection step-by-step. In the first step we select the optimal dimensionality. Therefore, a profile structure of the order 3, that is all main effects plus all two- and three-way associations, is used to define the design matrix $\bm{Z}$ ($Q = 25$), and a saturated structure (i.e., order 5) for the design matrix $\bm{W}$ ($T = 31$). Finally, $\bm{X}$ contains the main effects of all covariates (control variables) and the five personality variables ($P = 8$). We vary the dimensionality from 1 to 5. Table \ref{tab:nesda} shows that the two-dimensional solution has the lowest AIC. 

In the second step, we selected the profile structure ($\bm{Z}$), where we first compare the model with all main effects and two- and three-way association (order is 3) to the model with the model with all main effects and two-way associations (order is 2, $Q = 15$), and the model with only all main effects (order is 1, $Q = 5$). The latter has the lowest AIC. Than we checked whether any of the two-way associations among responses improves model fit over this model with only main effects of the responses. Therefore, we added interaction effects to $\bm{Z}$ ($Q$ becomes  6). Adding the two-way association between dysthymia and generalized anxiety disorders and between major depressive disorder and generalized anxiety disorder resulted in a lower AIC. Adding both, however, lead to an increase in AIC, so therefore we only include the association between dysthymia and generalized anxiety disorders. 

In the third step, we checked whether a personality characteristic can be left out of the model by deleting main effects in $\bm{X}$. We found that both Openness and Agreeableness can be left out of the model (so, final $P$ equals 6, three main effects of the control variables and the effects of neuroticism, extraversion, and conscientiousness). 

In the final, fourth step, we verified which associations among the responses are needed. Effects from the design matrix $\bm{W}$ are deleted, so $T$ becomes smaller. We see in Table \ref{tab:nesda} that all four and three-way associations do not contribute significantly to the model fit. The final design matrix $\bm{W}$ has main effects and all pairwise associations, so $T = 15$. 

The final model uses the main effects of the control variables (age, gender and education) plus main effects of neuroticism (N), extraversion (E), and conscientiousness (C) in the design matrix $\bm{X}$. The design matrix $\bm{Z}$ has a main effect for each response plus an association between dysthymia and generalized anxiety disorder (D:G). The design matrix $\bm{W}$ includes all main effects and two-way associations of the response variables. We will interpret this final model.  

As discussed in Sections \ref{sec:logodds} and \ref{sec:logoddsratio}, the main interest lies in the effects of the predictors on the underlying responses and their associations. Equations (\ref{eq:logodds}) and (\ref{eq:logoddsratio}) show the interpretational rules, where there is first a contrast of intercepts and thereafter the effect of predictors. Note that for the responses major depressive disorder (M), social phobia (S), and panic disorder (P) there is a main effect structure, while for the response variables dysthymia (D) and generalized anxiety disorder (G) there is a main and association effect structure.

The estimated coefficients are shown in shown in Table \ref{tab:nesda_coef}. The first row, indicated by ``1'', shows the contrast of estimated intercepts, whereas the remaining rows display the coefficients for the effect of the predictor variables on the (association among) responses, that is $\hat{\bm{B}}_x \hat{\bm{B}}'_z$.


The first row of Table \ref{tab:nesda_coef},  indicated by ``1'', shows the contrast of estimated intercepts ($m_k - m_l$) as shown in Equation (\ref{eq:logodds}) for the five response variables and the contrast $m_k + m_o - m_l - m_n$ as shown in Equation (\ref{eq:logoddsratio}) for the association between dysthymia and generalized anxiety disorder. 
This effect pertains to the estimated log odds (first five columns) or estimated log odds ratio (last column) when the predictor variables are null.
So, when the predictor variables equal zero, the estimated log odds of major depressive disorder (M) equals 1.19, whereas the estimated log odds ratio for dysthymia and generalized anxiety disorder (D:G) equals 1.01. 

The other rows, show the effect of a predictor variable on these log odds and log odds ratio. For the columns pertaining to major depressive disorder, social phobia and panic disorders, these effects can be interpreted similar to the coefficients of binary logistic regression. We see, for example, that with every standard deviation increase of neuroticism (N) the log odds of major depressive disorder (M) goes up by 0.53, the log odds of social phobia (S) goes up by 0.55, and the log odds of panic disorder (P) goes up by 0.17. So, when a person has a higher score on neuroticism the probability for each of these three disorders goes up. In a similar way, we can conclude that with larger values of extraversion the probabilities for having major depressive disorder or social phobia are smaller (coefficients are negative, i.e., -0.25 and -0.31), but the probability of having a panic disorder is higher (coefficient 0.11). 

As noted in Section \ref{sec:logodds}, when there is an association structure among the responses, such as in our final model among dysthymia and generalized anxiety disorder, we should refrain from interpreting the main effects as it has no simple structure anymore. Instead the interpretation should focus on the highest order effect. This effect is shown in the last column of Table \ref{tab:nesda_coef}, where we see that the log odds ratio between dysthymia and generalized anxiety disorder goes down by 0.38 with every standard deviation increase in neuroticism and goes up by 0.16 for every standard deviation increase in extraversion (E). With the canonical decomposition model, we can get estimates of these effects of predictors on the associations among response variables. 

\begin{table}[t]
\centering
\begin{tabular}{l|rrrrrr}
  \hline
 & D & M & G & S & P & D:G \\ 
  \hline
  1 & -1.61 & 1.19 & -0.82 & -0.48 & 0.03 & 1.01 \\ 
  Gender & -0.07 & 0.02 & -0.01 & -0.04 & 0.20 & -0.02 \\ 
  Age & 0.04 & 0.05 & 0.02 & 0.05 & 0.00 & -0.03 \\ 
  Edu & -0.04 & -0.22 & -0.05 & -0.16 & -0.32 & 0.17 \\ 
  N & 0.36 & 0.53 & 0.18 & 0.55 & 0.17 & -0.38 \\ 
  E & -0.25 & -0.25 & -0.10 & -0.31 & 0.11 & 0.16 \\ 
  C & -0.10 & 0.01 & -0.01 & -0.06 & 0.24 & -0.02 \\ \hline
\end{tabular}
\vspace{0.7cm}
\caption{Estimated implied coefficients ($\bm{B}_x \bm{B}'_z$) of predictors (rows) on the responses (columns) and the association between two responses (D:G) for the NESDA data. Abbreviations for the predictor variables: Edu = number of years of education, N = Neuroticism, E= Extraversion, C = Conscientiousness. Abbreviations for the response variables: D = dysthymia, M = major depressive disorder, G = generalized anxiety disorder, S = social phobia, P = panic disorder. }
\label{tab:nesda_coef}
\end{table}

Finally, we inspect the associations among responses not influenced by the predictor variables, that are, the associations defined in $\bm{W}$ but not in $\bm{Z}$. The effects correspond to the contrasts of intercepts,  as in the first terms in Equation (\ref{eq:logodds}) and (\ref{eq:logoddsratio}). These associations are not influenced by the predictors. The estimated pairwise associations among the response variables that are not influenced by the predictor variables are 1.05 for DM (i.e., the association between dysthymia (D) and major depressive disorder (M)), -0.01 for DS, 0.33 for DP, 0.61 for MG, -0.14 for MS, 0.16 for MP, 0.45 for GS, 0.62 for GP, and 1.23 for SP. This means that the association between dysthymia and major depressive disorder does not depend on the predictor variables, it is constant with an odds ratio of $\exp(1.05) = 2.86$. We see that there are some considerable dependencies among the responses that are not modeled by the use of a lower dimensionality or explained by the predictors. We also see that dysthymia (D) and social phobia (S) are conditionally independent as the log odds ratio is almost 0. 

These data could have been analysed using a (reduced rank) multinomial logistic model. In such an analysis, we would use the $2^5 = 32$ response profiles as the categories of a nominal response variables. Age, gender, education, and the five personality characteristics would be used as predictor variables. In such a model, we would obtain effects of the predictor variables on the log odds of one profile class against another. This would lead to a large matrix of coefficients (i.e., the dimension of this matrix is $P$ by $2^R - 1$), each referring to a change in the log odds corresponding to a unit change in the predictor. However, the effect of a predictor, say neuroticism, on the underlying response variables, such as major depressive disorder, can not be obtained from such an application. Using our canonical decomposition model, we can obtain such effects.  


\section{Discussion}

In this paper, we described a decomposition with linear constraints of the canonical parameters of a multinomial model. This decomposition is especially useful when the number of categories of the variable of interest is large and when there is external information available for those categories. One specific type of data we emphasized, is when the categorical variable defines a profile of underlying binary (or categorical) variables. The external information is then implicit in the formation of the profiles. The canonical decomposition allows to estimate effects of predictor variables on the underlying binary response variables, but also to estimate effects of predictors on the association between the underlying response variables. We derived equations to describe changes in 
\begin{enumerate}
    \item log odds of a response variable given a change in the predictor variables; 
    \item log odds ratios of pairs of response variables given a change in the predictor variable. 
\end{enumerate}
Similar formulas could be derived for higher order associations. 

The canonical decomposition model is a full maximum likelihood method, assuming a multinomial distribution for the response profiles. We developed an MM algorithm for maximum likelihood estimation of the model parameters. In MM algorithms, a majorization function is defined in every iteration which is subsequently minimized. MM algorithms generally have a linear rate of convergence \citep{hunter2004tutorial}, that is, they often need many iterations to converge. On the other hand, the computations within the iterations are usually simple. In our case, the majorization function turns out to be a least squares function that can be solved using a singular value decomposition. Alternatively, the two matrices of coefficients can be updated in an alternating fashion. We like to point out that, in comparison to IRLS algorithms that have been proposed in the literature for fitting reduced rank models \citep{yee2003reduced, yee2015}, there is no need to invert a matrix in each and every iteration. The derivation of the MM algorithm might look a bit more complex, but the implementation is simpler. MM algorithms often require a few more iterations than IRLS algorithms \citep{derooij2023new}. \cite{heiser1995convergent} discussed ways to decrease the number of iterations for MM algorithms.  

When applied to profiles of multivariate binary response variables, the canonical decomposition model can be understood as a new regression method for such data that can handle both numerical and categorical predictors. When all predictor variables are dichotomous or categorical the canonical decomposition model in maximum dimensionality becomes equivalent to some loglinear models. We discussed the relationships in detail in Section \ref{sec:complla} and illustrated the equivalences in Section \ref{sec:ACMdata}. For purely categorical variables, the loglinear approach is more general than the multinomial canonical decomposition. Therefore, we conclude that in this case (i.e., when all variables are categorical) a researcher better uses loglinear models for the statistical analysis.

However, in loglinear analysis all variables should be categorical. Our canonical decomposition allows also for numerical predictors, as we showed in Section \ref{sec:nesda}. In this analysis there are five dichotomous response variables and eight predictor variables of which seven are numeric. We showed in detail how the final model can be interpreted. 

For these data also a (reduced rank) multinomial logistic model could be applied. Such an application would lead to a predictive relationship between the predictors (neuroticism, extraversion, etc.) and the responses profile classes, but no insight into the effects of the responses on the binary response variables. These latter effects can be obtained from our new modeling approach.

We focused on binary response variables, but when we have profiles of multi-category response variables (nominal or ordinal), the canonical decomposition model can simply be adapted. Only the definition of the design matrices $\bm{Z}$ and $\bm{W}$ needs to be changed. As these follow design matrices of loglinear models, such an adaption should not be too difficult. Also the interpretational rules, as we derived in Section \ref{sec:dichotomous}, can be simply adapted, and will still be in the form of changes in log odds or log odds ratios corresponding to changes in the predictor variables. 

Another occasion when the canonical decomposition model might be favorable is when the contingency table is sparse. In loglinear models, maximum likelihood estimates do not exist when a count in the sufficient statistics is zero \citep[][Section 9.8.2]{agresti2002categorical}. Using the canonical decomposition model, by reducing the dimensionality we might obtain estimates of effects that do not exist in loglinear models. This hypothesis follows from models for two-way contingency tables with one or more observed zero counts, such that the parameters of a saturated loglinear model become infinite or may not exist. Applying a reduced rank model, such as the RC-association model \citep[][Section 9.6]{goodman1979simple, agresti2013categorical}, gives valid parameter estimates. Further research is needed to investigate canonical decomposition models for sparse data.  

In Section \ref{sec:relations}, we showed close connections to double constrained correspondence analysis, a method often used in ecology. In double constrained correspondence analysis, a primary interest is the fourth corner correlation. This correlation describes, in our terminology, the association between the predictor variables and the characteristics of the response categories. In that sense, our formulas for changes in estimated log odds given a change in the predictor variables, as derived in Section \ref{sec:dichotomous}, resemble those fourth corner correlations. 

Doubly constrained correspondence analysis usually results in a biplot representation of the data \citep{terbraak2018algorithms}. Biplots have been first described by \cite{gabriel1971biplot}, and further worked out in \cite{gower1996biplots} and \cite{gower2011understanding}. The results of the multinomial canonical decomposition model can probably also be represented using a biplot. Predictor variables and response variables can be represented by arrows or variable axes corresponding to the estimated effects $\hat{\bm{B}}_x$ and $\hat{\bm{B}}_z$. Further research is needed to work out, for example, the valued markers on such variable axes and whether the estimated $\hat{\bm{b}}_w$ can be included in such a biplot.

As discussed, the multinomial canonical decomposition model allows for both categorical and numeric variables. There have been other attempts to multivariate classification using mixtures of dichotomous and continuous predictors. The location model \citep{olkin1961multivariate, afifi1969multivariate, krzanowski1975discrimination, krzanowski1976canonical, krzanowski1993location} is such an approach. In the location model, the joint distribution of the variables is decomposed into a marginal distribution of the dichotomous variables, and a conditional (multivariate) normal distribution of the continuous variables. \cite{krzanowski1976canonical} shows in detail how to use the location model for classification.  The location model is a \emph{generative} model whereas the model we propose in this paper is a \emph{discriminative} one \citep[cf.][]{ng2001discriminative}. Each of these two paradigms has its advantages and disadvantages \citep{efron1975efficiency, rubinstein1997discriminative}, depending on the circumstances. One advantage of our model is that it needs no assumption about the distribution of the continuous variables, whereas the location model assumes these variables to follow a multivariate normal distribution.

The canonical decomposition model we introduced is a so-called conditional model, that is, the effects on response variable $r$ should be interpreted conditional on the other response variables, similar as the interpretation in loglinear models. This type of model can be contrasted with marginal models \citep{McCullagh1989GLM, liang1992multivariate, glonek1995multivariate} where the effects of predictors on the responses are marginal effects. Maximum likelihood estimation of marginal models is, however, difficult because the likelihood is based on the joint probabilities whereas the model only considers marginal probabilities (\citep{agresti2013categorical}. With only categorical variables an algorithm has been proposed by \cite{bergsma2009marginal}. With numerical predictors and having more than two response variables researchers commonly use generalized estimating equations \citep{liang1986longitudinal, liang1992multivariate, hardin2002generalized}, but that is not a maximum likelihood method. 
A marginal approach can also be specified by reduced rank logistic models \citep{yee2003reduced, derooij2023new}, where conditional on the reduced rank the responses are assumed to be independent. Such an approach, however, does not allow to also model the relationships between responses. 

For model selection, we proposed to use a step-wise approach based on the AIC. For the number of parameters we used
$\mathrm{npar} = T + S(P + Q - S)$, which is a naive estimator \citep{mukherjee2015degrees}. For Gaussian reduced rank models there are unbiased estimators for the effective number of parameters \citep{mukherjee2015degrees}, but these have not been generalized for other reduced rank models. \cite{mukherjee2015degrees} point out that having a better estimator of the effective degrees of freedom influences the final model choice. Further research is needed on this topic. 
Following \cite{yu2013model}, we suggested a step-wise model selection approach, starting with selection of the optimal dimensionality, then the selection of the profile score structure, followed by the selection of the object score structure (which predictor variables), and finally determining the association structure among the response variables. Such a step-wise approach facilitates model selection, but might not be optimal in all circumstances. Further research is needed on this topic. 

To conclude this paper, a word on software. We developed R-functions for estimation of the model parameters using the updating scheme of Section \ref{sec:updates}. The functions are made public on the \texttt{github}-page of the author and will be included in the R-package \texttt{lmap} \citep{lmappackage}.

\section*{Statements and declarations}

\paragraph*{Data availability:} This paper uses two empirical data sets as illustrations. 
\begin{enumerate} 
\item The alcohol, cigarette, and marijuana data are available in the R-package \texttt{CatDataAnalysis}
\item The depressive and anxiety disorder data set is available upon request from the nesda consortium, see https://www.nesda.nl/nesda-english/
\end{enumerate}
R-code for the analyses of these data sets is available on the \texttt{github}-page of the author. 

\paragraph*{Funding:}
The author declares that no funds, grants, or other support were received during the preparation of this manuscript. The manuscript was revised while  the author was a fellow at the Netherlands Netherlands Institute for Advanced Study in the Humanities and Social Sciences (NIAS) in Amsterdam.

\paragraph*{Competing interests:}
The authors have no relevant financial or non-financial interests to disclose.

\paragraph*{Author contributions:}
As this is a single author paper, everything was done by the main author.

\bibliographystyle{apalike}
\bibliography{melodic.bib}  

\end{document}